\documentstyle[pra,epsf,aps,twocolumn]{revtex}
\begin{document}
\title{Minimal Model for Dilatonic Gravity and Cosmological Constant}
\author{P.~P.~Fiziev \thanks{ E-mail:\,\, fiziev@phys.uni-sofia.bg}}
\address{Department of Theoretical Physics, Faculty of Physics,
Sofia University, 5 James Bourchier Boulevard, Sofia~1164,
Bulgaria
 }
\maketitle
\begin{abstract}
We consider: minimal scalar-tensor model of gravity with
Brans-Dicke factor $\omega(\Phi)\equiv 0$ and cosmological factor
$\Pi(\Phi)$; restrictions on it from gravitational experiments;
qualitative analysis of new approach to cosmological constant
problem based on the huge amount of action in Universe;
determination of $\Pi(\Phi)$ using time evolution of scale factor
of Universe.

\noindent{PACS number(s): 04.50.+h, 04.40.Nr, 04.62.+v}
\end{abstract}
%%%%%%%%%%%%%%%%%%%%%%%%%%%%%%%%%%%%%%%%%%%%%%%%%%%%%%%%%%%%%%%%%%%
%\draft
\sloppy
%\scrollmode
%%%%%%%%%%%%%%%%%%%
\newcommand{\lfrac}[2]{{#1}/{#2}}
\newcommand{\sfrac}[2]{{\small \hbox{${\frac {#1} {#2}}$}}}
\newcommand{\ben}{\begin{eqnarray}}
\newcommand{\een}{\end{eqnarray}}
\newcommand{\la}{\label}
%
%%%%%%%%%%%%%%%%%%
\section{Introduction}

The recent astrophysical observations of the type Ia supernovae \cite{Ia},
CMB \cite{CMB}, gravitational lensing  and galaxies cluster's
dynamics (see the review articles \cite{CosmTri} and the references therein)
gave us a strong and {\em independent}
indications of existence of some new kind of energy in the Universe,
needed to explain its accelerated expansion.
Nevertheless we are not complete confident in these preliminary
results, combining them with old problems of cosmology and astrophysics,
we can conclude that most likely some further generalization and enlargement
of the frameworks of the well established fundamental laws of physics,
and in particular, of the laws of gravity, is needed.

At present, general relativity (GR) is the most  successful theory of
gravity at scales of laboratory-, Earth surface-, solar system- and
star systems.  It gives a quite good description of gravitational
phenomena in the galaxies and at the scales of the whole Universe \cite{WW}.
But without some essential changes it is problematic in explanation
of the rotation of the galaxies and their
motion in the galactic clusters,
the initial singularity problem,
the physics in the early
Universe and the inflation \cite{inflation},
the present days accelerated expansion
of the Universe \cite{Ia}-\cite{CosmTri}
and the famous vacuum energy problem
\cite{Weinberg}.

The most promising modern theories of gravity, like
supergravity and (super)string theories \cite{Strings},
having a deep theoretical basis, incorporate naturally GR.
Unfortunately, at least at present, they are not developed enough to
allow {\em a real} experimental test,
and introduce a large number of new fields
without any {\em direct experimental evidences} for doing this.

Therefore it seems meaningful to look for some {\em minimal
extension} of GR which is compatible with known gravitational
experiments, promises to overcome at least some of the above problems
and may be considered as a necessary part of more general modern theories.

It is most likely that such  minimal extension must include
one new scalar-field degree of freedom.
Its contribution to the action of the theory
can be described in different (sometimes equivalent) ways, being
not fixed a priori and there exist many attempts in this direction,
starting from Jordan-Fierz-Brans-Dicke theory of variable gravitational
constant and its further generalizations -- the so called
scalar tensor theories of gravity \cite{BD}
which have been proved to be the most natural extension of GR \cite{Damoor+}.
Different models of this type ware used in the inflationary scenario
\cite{inflation} and in the more recent quintessence models \cite{Q}.
For the latest development of the scalar-tensor theories with respect to
accelerated expansion of the Universe one can consult
the recent article \cite{E-F_P}.

One more model called {\em a minimal dilatonic gravity} (MDG) was
proposed in the article \cite{F9911037}. Being a model with one
additional scalar field $\Phi$ which couples non-minimally with
the space-time metric, it differs from the known inflationary
models with spin zero inflation field. At the same time being
mathematically equivalent to some of the quintessence models, MDG
describes {\em a complete different physics} (see \cite{E-F_P}),
because of the relatively big mass $m_\Phi \geq 10^{-4} eV$ of the
scalar field $\Phi$ \cite{F9911037}. One has to remind the reader
that in the standard quintessence models two different
possibilities ware used until now (see \cite{St1} and the
references therein): a scalar field with a typically extremely
small mass $\sim 10^{-33} eV$, or massless scalar field with
inverse-degree-potential \cite{Q}. In both cases these models
suffer of some difficulties \cite{Q1}.

The dilatonic-gravity action in the MDG model has the form
\ben
{\cal A}_{G,\Lambda}=-{\sfrac c {2\bar\kappa}}\int d^4 x\sqrt{|g|}
\Phi \bigl( R + 2 \Lambda \Pi(\Phi) \bigr)
\la{A_Gc}
\een
and corresponds to the specific choice $\omega(\Phi)\!\equiv\!0$ of the
Brans-Dicke parameter.
In the Eq. (\ref{A_Gc}) $\Lambda$ is the cosmological constant and the function
$\Pi(\Phi)$ presents a dimensionless cosmological factor.

The matter action ${\cal A}_M$ and matter equations of motion are supposed
to have usual GR form and do not include the scalar field $\Phi$.
This is our {\em most important physical assumption}
and it means that the dilaton field $\Phi$
{\em does not interact directly} with the usual matter of any kind.
Its influence on this matter is indirect -- only due to the interaction
of  dilaton field $\Phi$ with the space-time metric.

Equations for metric $g_{\alpha\beta}$ and dilaton field $\Phi$:
\ben
\Phi \left(G_{\alpha\beta}\!-\!\Lambda \Pi(\Phi)
g_{\alpha\beta}\right) \!-\!(\nabla_\alpha
\nabla_\beta\!-\!g_{\alpha\beta}{\Box})\Phi \!=\! {\sfrac
{\bar\kappa} {c^2}} T_{\alpha\beta}, \nonumber \\
{\Box}\Phi\!+\!\Lambda {\sfrac {dV}{d\Phi}}(\Phi)=\! {\sfrac
{\bar\kappa} {3 c^2}}T
\la{FEq}
\een
yield usual energy-momentum conservation law
\ben
\nabla_\alpha\,T^\alpha_\beta=0
\la{ConsLow}
\een
and the important relation:
\ben
R+2\Lambda {\sfrac{dU}{d\Phi}}(\Phi)=0.
\la{RV}
\een
The quantities ${\sfrac {dV}{d\Phi}}:={\sfrac 2 3}
\Phi\left(\Phi{\sfrac{d\Pi}{d\Phi}}-\Pi\right)$ and $U(\Phi):=\Phi
\Pi(\Phi)$ introduce {\em dilatonic potential} $V(\Phi)$ -- in Eq.(\ref{FEq})
and  {\em cosmological potential}
$U(\Phi)$ -- in Eq.(\ref{RV}).

It is remarkable that introducing the dilaton field $\Phi$
in MDG we do not need to prescribe some new charges,
or other novel properties to the usual matter.
As seen from Eq. (\ref{FEq}), this field is complete determined
by the trace of the energy-momentum tensor of matter.

One has to stress that because of the condition $\omega(\Phi)\!\equiv\!0$
the existence of nontrivial relativistic dynamics
and propagation of the dilaton field $\Phi$ in vacuum
is deeply connected with {\em the nonzero space-time curvature}.
Technicaly, the second order dynamical equation for $\Phi$
in the system (\ref{FEq}) is obtained by
contraction of the generalized Einstein equations
and making use of the algebraic relation (\ref{RV}),
derived by variation of the action
(\ref{A_Gc}) with respect to the dilaton field.
The second order derivatives of the field $\Phi$ are created
during the two-fold integration by parts of the corresponding
terms in variation of the specific action (\ref{A_Gc})
with respect to the metric, not with respect to the very field $\Phi$.
Therefore, in a flat space-time, the zero scalar curvature $R$ in action
(\ref{A_Gc}) would  lied to a non-existence of
dynamics of the dilaton field.
This simple argument, {\em being specific for MDG},
shows a deep connection of the field $\Phi$
with the space-time curvature, i.e., with the gravity.
Therefore it seems natural to treat the dilaton field $\Phi$
as a "scalar part of gravity",
instead of considering it as a new kind of matter scalar field.

The action (\ref{A_Gc}) can be considered as a Helmholz action of
nonlinear gravity (see \cite{NLG} and the references therein)
with lagrangian $L_{NLG} \sim \sqrt{|g|} f(r)$,
$r=R/\Lambda$ being dimensionless scalar curvature \cite{F9911037}.
Indeed, because of the absence of a
Brans-Dicke kinetic term $\sim\omega(\Phi)(\nabla
\Phi)^2$, the variation of the action (\ref{A_Gc}) with
respect to the dilaton field $\Phi$  gives the local {\em
algebraic} relation (\ref{RV}) between scalar curvature $R$
and dilaton $\Phi$, instead of differential equation.
If ${\sfrac{d^2}{d\phi^2}}U(\Phi)\,\, {/ \hskip
-.33cm\equiv}\, 0$, one can solve the relation (\ref{RV}) with
respect to the field $\Phi$ and this field becomes a {\em local} function
of the scalar curvature: $\Phi= \Phi(r)$. (For $\omega\neq 0$ the
last relation would have a non-local integral form.) The substitution
of this function back into the action (\ref{A_Gc}) transforms it
to the action of nonlinear theories of gravity ${\cal A}_{NLG}=
-{\sfrac {c\Lambda} {2\bar\kappa}}\int d^4 x \sqrt{|g|}\,f(r)$ with
$f(r) = r\Phi(r) + 2 U(\Phi(r))$.
The inverse correspondence -- from NLG to MDG --
can be described in a simple way, too.
For a given function $f(r)$ we have to solve the algebraic
equation ${\sfrac d {dr}}f(r) = \Phi$ with respect to the variable $r$.
This gives a function
$r=r(\Phi)$. Then $U(\Phi)= - \int r(\Phi) d\Phi + const $.

It is well known that the nonlinear gravity can be created by
the quantum corrections which appear after quantization of the
classical fields in curved space-time \cite{QNLG}.
This was the physical basis of the original Starobinsky
model of inflation \cite{Starobinsky80}. The modern development of
this model one can find in the recent articles \cite{Hawking2000}.

The total energy momentum of the vacuum fluctuations in this model
cannot be obtained by varying a local action, see the article by
A.~Vilenkin in \cite{Starobinsky80}. Therefore, in the general
case  MDG differs both physically and mathematically from the
Starobinsky model and its modern developments. For example, the
cosmological perturbations in the two models are essentially
different. Nevertheless, it turns out that under proper particular
choice of the cosmological factor $\Pi(\Phi)$, MDG coincides with
Starobinsky model in the case of {\em the conformally flat}
Robertson-Walker (RW) metric. This happens just because the term
which yields the essential difference between the two models is
proportional to the Weyl conformal curvature tensor, but this
curvature is zero in the conformally flat spaces with RW metric
\cite{H-J_S}.

The last consideration indicates that probably we have to look
for the roots of our MDG in quantum field theory in curved space-time.

At present, a well known candidate for a self-consistent quantum
theory of gravity is the superstring theory. It turns out that the
MDG can be considered as a four dimensional version of the low
energy limit (LEL) of the string theory in some {\em new frame}.
In this new frame the action (\ref{A_Gc}) is precisely the stringy
three-level effective action {\em for the metric and the dilaton
only}, corresponding to the lowest order of string loop expansion,
i.e., for the most constant part of the string theory which has
the same form in all string models.

Indeed, let us consider the $D$-dimensional LEL lagrangian
in stringy frame  (SF) \cite{Strings}:
$${}_{{}_S}\!L_{{}_{LEL}} \sim \sqrt{|{}_{{}_S}\!g|}
e^{-2\phi}\left({}_{{}_S}\!R +
4 {}_{{}_S}\!(\partial\phi)^2 +{}_{{}_S}\!V_{SUSY}(\phi)\right).$$
After Weyl conformal transformation:
$g_{\mu\nu} \rightarrow e^{b\phi}g_{\mu\nu}$ to some new conformal
frame, which depends on the parameter $b$, it acquires the form:
\ben
L_{{}_{LEL}} \sim
\sqrt{|g|}e^{p_1(b)\phi}
\left(R\!+\!p_2(b)
(\partial\phi)^2\!+\!V_{SUSY}(\phi)\right)
\nonumber
\een
where $p_1(b)=(D-2)b-2$ and
$p_2(b)=4\!-\!4(D\!-\!1)b\!+\!(D\!-\!1)(D\!-\!2) b^2$ are polynomials in $b$
of degree 1 and 2, correspondingly.

Now, if we chose the parameter $b=b_{{}_E}={2\over {D-2}}$, i.e., if
$p_1(b_{{}_E})=0$, we will obtain the well known Einstein frame (EF) in which
\ben
{}_{{}_E}\!L_{{}_{LEL}} \sim
\sqrt{|{}_{{}_E}\!g|}
\left({}_{{}_E}\!R\!+\!{4\over {D-2}}\,
{}_{{}_E}\!(\partial\phi)^2\!+\!{}_{{}_E}\!V_{SUSY}(\phi)\right).
\la{L_E}
\een

But there exists another simple choice: $p_2(b_{{}_F})=0$ which
gives $b=b_{{}_F}={2\over {D-2}}\left(1 \pm {1\over
{\sqrt{D-1}}}\right)$ and a new conformal frame which we call {\em
a fundamental frame} (FF). If one sets $\Phi := exp\left(\pm
{{2\phi}\over {\sqrt{D-1}}}\right)$, one reaches the following
simple form of string LEL lagrangian for metric and dilaton fields
in FF: \ben {}_{{}_F}\!L_{{}_{LEL}} \sim
\sqrt{|{}_{{}_F}\!g|}\,\bigl( \Phi
{}_{{}_F}\!R\!+\!{}_{{}_F}\!U_{SUSY}(\Phi)\bigr). \la{L_F} \een
Obviously, under proper re-interpretation of all terms in it, this
lagrangian gives the action (\ref{A_Gc}) as a truncated
4-dimensional LEL action in graviton-dilaton sector of the string
theory, i.e., as an action, obtained neglecting i)the contribution
of all other string excitations, ii)the contribution of the
fields, connected with the higher dimensions and iii)contribution
of higher order terms of string perturbation theory.

At present the exact form of the potential
${}_{{}_S}V_{SUSY}(\phi)$ is not known. It may originate from SUSY
breaking due to gaugino condensation \cite{Strings}, or may appear
in the theory in some more sophisticated way.

The form of lagrangian (\ref{L_F}) in FF is more
simple than the corresponding form in EF -- (\ref{L_E}).
In the FF we have a more direct interpretation of the MDG
{\em in the spirit} of Jordan-Fierz-Brans-Dicke
theory, i.e. as a theory with variable gravitational "constant"
$G=\bar G \Phi^{-1}= \bar G exp\left(\mp {{2\phi}\over {\sqrt{D-1}}}\right)$.
Because of the quadratic equation $p_2(b_{{}_F})=0$
we have two possible values of the
parameter $b_{{}_F}$ and two fundamental frames: $FF_\pm$.
In each of them the gravitational "constant"
depends exponentially on the original (i.e., defined in the SF)
dilaton field $\phi$, but the signs of the arguments
in the corresponding exponents are opposite.

Note that using other values of the parameter $b$ in the
exponential factor $e^{b\phi}$ of the Weyl conformal
transformation, one can produce a kinetic term for the dilaton in
the action of theory with any desired value of the coefficient
$\omega$. It is clear that among the infinitely many possible
conformal frames the above two: the Einstein frame and the
fundamental frame are distinguished ones.

Usually one prefers to work in Einstein frame, because in it the
field variables ${}_{{}_E}\!g_{\mu\nu}$ and $\phi$ are separated
and the corresponding Cauchi problem is well posed \cite{E-F_P}.
Thus, the choice of Einstein frame is a convenient mathematical
tool which is analogous to the choice of normal coordinates in
usual mechanics and field theory. In the reference \cite{E-F_P} an
additional description of this property of Einstein frame is
stressed: in it we have no mathematical mixing between "true"
helicity-0 excitation $\phi$ and helicity-2 excitation
${}_{{}_E}g_{\mu\nu}$, in contrast to the situation in other
conformal frames. For example, in Jordan frame for scalar-tensor
theories of general type these excitation are mixed, see for
details the reference \cite{E-F_P}.

It is well known that the EF is not the physical one and
one needs to find the physical frame to reach a right interpretation
of the results, see for example \cite{E-F_P} and the references therein.

Im MDG the true  separation of the {\em physical} properties
of the helicity-0  and helicity-1 degrees of freedom
takes place in the fundamental frame:

1) Because of the condition $\omega(\Phi)\equiv 0$ in MDG we have a local
functional dependence between the fields $\phi$ and $\Phi := exp\left(\pm
{{2\phi}\over {\sqrt{D-1}}}\right)$,
instead of the differential equation (2.4b) in the reference \cite{E-F_P}.
Therefore the field $\Phi$ is not {\em physically}
different from the field $\phi$, at least locally, and
it carries all physical properties
of the true helicity-0 degree of freedom,
nevertheless these properties are described
in a different mathematical way.

2) In contrast to other frames in FF
the helicity-0 field degree of freedom does not interact directly
with matter at all,
and only the helicity-2 field degree of freedom is responsible
for the interaction of gravity with matter.

Some time ago, the FF was recognized to be an useful tool
in the two-dimensional models of dilatonic gravity,
both classical and quantum ones --
see the recent article \cite{Cavaglia} and the references therein.
Moreover, for the exact quantization of all $D=2$ models of pure dilatonic
gravity {\em with arbitrary potential} $U(\Phi)$
it turned out to be critical to work just in the FF frame \cite{MGK}.

Now our main physical assumption may be formulated as a hypothesis
that {\em the week equivalence principle is valid precisely in
FF}. This means that just in FF we are to set the action for usual
matter in its GR form, i.e., we accept the FF as a physical frame
in which the physical observations and experiments are performed.
They are described in the terms of standard non-Euclidean
4-space-time geometry. In other words we assume that in these
experiments one is testing just the FF geometry using {\em the
real physical objects}. Hence the name "fundamental frame".

The string theory is supposed to work in the well studied and
relatively simple way at Planck scales, but in it one needs some
additional (at present unknown) procedures for describing the
usual matter. As a result, at present we do not know how to treat
the real matter (build of electrons, protons, $\pi$-mesons,
e.t.c.) in the framework of string theory. In particular, the
interaction of stringy dilaton with usual matter is unknown, see
the discussion of this problem and possible violation of the week
equivalence principle due to the interaction of dilaton with usual
matter in \cite{Damour_Polyakov}. In this situation our choice of
FF as a physical frame justifies the very string theory in the
spirit of Fierz article \cite{Fierz}, where he first noticed that
the extremely high precision of the week equivalence principle
(nowadays it is at the level of $10^{-12}$) suggests that the
coupling of matter and gravity must have {\em an exact metric
form}, but there still exist an open possibilities to change the
Einstein-Hilbert action of GR, see the recent reference
\cite{DamourExp}, too. If successful, our model of MDG can help
the further development of string theory as a possible physical
description of the real world.

The action (\ref{A_Gc}) appears, too, in a new model of gravity
with torsion and unusual local conformal symmetry after its
breaking in metric-dilaton sector \cite{PF1}
and in $D=5$ Kaluza-Klein theories \cite{Fujii}.

For boson stars the MDG was tested  in \cite{PF2}. There it was
shown that the star structure is slightly sensitive mainly to the
mass term in the dilatonic potential (typically into a few
percent) and do not depend on the exact form of this potential.

Investigation of MDG was started by O'Hanlon  in connection with
Fujii's theory of massive dilaton \cite{Hanlon}, but {\em without}
any relation with cosmological constant problem and other problems
of cosmology and astrophysics.

An {\em essential new element} of our MDG is the {\em
nonzero cosmological constant $\Lambda$} \cite{F9911037}.
Nevertheless at present still exist doubts in astrophysical data:
$\Omega_{\Lambda}\!=\!.65\pm .13,\,\,\,H_0\!=\!(65\pm
5)\,km\,s^{-1}Mps^{-1}$  which determine
$$\Lambda^{obs}=3\Omega_\Lambda H_0^2 c^{-2}= (.98 \pm .34) \times
10^{-56} cm^{-2}$$ we accept this observed value of cosmological
constant {\em as a basic quantity} which  {\em defines natural
units} for all other cosmological quantities, namely: cosmological
length: $A_c := 1/\sqrt{\Lambda^{obs}} = (1.02\pm .18)\times
10^{28} cm$, cosmological time: $T_c:=A_c/c= (3.4\pm .6)\times
10^{17} s=(10.8 \pm 1.9) Gyr$, cosmological energy density:
$\varepsilon_c:={\frac {\Lambda c^2} {\kappa}}= (1.16\pm
.41)\times 10^{-7} g~cm^{-1}~s^{-2}$, cosmological energy: $E_c:=3
A_c^3\varepsilon_c= 3\Lambda^{-1/2} c^2 \kappa^{-1} =(3.7\pm .7)
\times 10^{77} erg$, cosmological momentum: $P_c:= 3c / (\kappa
\sqrt{\Lambda^{obs}}) = (1.2 \pm .2)\times 10^{67} g~cm~s^{-1}$
and {\em cosmological unit for action}: $${\cal A}_c:=
3c/(\kappa\Lambda^{obs})=(1.2 \pm .4)\times 10^{122}\,\hbar,$$
\noindent $\kappa$ being Einstein constant. Further we use
dimensionless variables like: $\tau := t/T_c$, $a := A/A_c$, $h:=
H\,T_c$ ($H:=A^{-1}dA/dt$ being Hubble parameter), $\epsilon_c :=
\varepsilon_c / |\varepsilon_c|=\pm 1$, $\epsilon:=\varepsilon /
|\varepsilon_c|$-matter energy density, etc.

In our special scalar-tensor model of gravity the cosmological
factor $\Pi(\Phi)$ is the {\em only unknown function} which has to
be chosen to comply with gravitational experiments and
observations and to solve the {\em inverse cosmological problem}
described in the last Section.

\section{Solar System and Earth-Surface Gravitational Experiments}
From known gravitational experiments one can derive the following
properties of cosmological factor $\Pi(\Phi)$:

1. MDG with $\Lambda=0$ contradicts to solar system gravitational
experiments. The cosmological term $\Lambda\Pi(\Phi) \neq 0$ in
action (\ref{A_Gc}) is needed to overcome this problem.

2. In contrast to O'Hanlon's model we wish MDG to reproduce GR
with $\Lambda\!\neq\!0$ for some
$\Phi\!=\!\bar\Phi\!=\!const\neq\!0,$ i.e., the original de Sitter
solution. Then we derive for cosmological factor of this solution
the conditions:
\ben\Pi(\bar\Phi)=1,\,\,\, {\sfrac
{d\Pi}{d\Phi}}(\bar\Phi)=\bar\Phi^{-1},\,\,\, {\sfrac
{d^2\Pi}{d\Phi^2}}(\bar\Phi)={\sfrac 3 2}p^{-2}\bar\Phi^{-2}
\la{CFC}\een \noindent as follows: i)From action (\ref{A_Gc}) we
obtain the first {\em normalization condition} and Einstein
constant $\kappa=\bar\kappa/\bar\Phi$; ii)In vacuum, far from
matter MDG have to allow {\em week field approximation}:
$\Phi=\bar\Phi(1+\zeta)$ ($|\zeta|\ll 1$). Then the {\em
linearized} dilaton equation (\ref{FEq}): $\Box \zeta+
\zeta/l_\Phi^2\!=\!{\sfrac {\kappa}{3 c^2}}T$ gives the second
condition and iii)Taylor series expansion of the function
${\sfrac{dV}{d\Phi}}(\Phi)$ around the value $\bar\Phi$ introduces
dimensionless Compton length of dilaton $p\!=\!{\frac
{l_\Phi}{A_c}}$ and gives the third of conditions (\ref{CFC}). As
a result we obtain
\ben\Pi\!=\!1\!+ \zeta\!+{\frac 3 {4 p^2
}}\zeta^2\!+O(\zeta^3).\la{CF}\een

3. {\em Point particles of masses $m_a$} as source of metric and
dilaton fields give in Newtonian approximation gravitational
potential $\varphi({\em r})$ and dilaton field $\Phi({\em r})$:
\ben \varphi({\em r})/c^2\!=\! - {\sfrac G {c^2}}\!\sum_a\!{\sfrac
{m_a}{|{\em r - r}_a|}}\! \left(\!1\!+\!\alpha(p) e^{-|{\em r -
r}_a|/l_\Phi} \right) \nonumber\\ - {\sfrac 1 6}p^2
\sum_a\!{\sfrac{m_a} M}\left(|{\em r - r}_a|/l_\Phi\right)^{\!2},
\een
\ben \!\Phi({\em r})/\bar\Phi\!=\!1+\!{\sfrac 2 3} {\sfrac
G{c^2(1-{\sfrac 4 3} p^2)}} \sum_a\!{\sfrac {m_a}{|{\em r -
r}_a|}} e^{- |{\em r - r}_a|/l_\Phi}, \la{SolNewton} \een
$G\!=\!{\sfrac {\kappa c^2} {8\pi}}(1\!-\!{\frac 4 3}p^2)$ is
Newton constant, $M\!=\!\sum_a m_a$. The term $-{\sfrac 1 6}p^2\!
\sum_a{\sfrac{m_a} M}\left(|{\em r\!-\!r}_a|/l_\Phi\right)^2\!=\!
-{\sfrac 1 6} \Lambda|{\em r}-\!\sum_a \!{\sfrac {m_a} M}{\em
r}_a|^2\!+\!const$ in $\varphi$ is known from GR with $\Lambda
\neq 0$. It represents an {\em universal anti-gravitational
interaction} of test mass $m$ with any mass $m_a$ via repellent
elastic force
\ben {\em F}_{\!{}_\Lambda\,a}={\sfrac 1 3}\Lambda m c^2
{\sfrac{m_a} M}({\em r - r}_a). \la{Fel} \een

For {\em solar system distances} $l\leq 1000 AU$ neglecting the
$\Lambda$ term ( of order $\leq 10^{-24}$) we compare the
gravitational potential $\varphi$ with specific MDG coefficient
$\alpha(p)={\frac{1+4p^2}{3-4 p^2}}$ with Cavendish type
experiments and obtain an experimental constraint $l_\Phi \leq
1.6$ [mm], or $p < 2\times 10^{-29}.$ Hence, in the solar system
the factor $e^{-l/l_\Phi }$ has fantastic small values
($<\exp(-10^{13})$ for the Earth-Sun distances $l$, or $<
\exp(-3\times 10^{10})$ for the Earth-Moon distances $l$) and {\em
there is no hope to find some differences between MDG and GR} in
this domain.

The corresponding constraint $m_\Phi c^2 \geq 10^{-4}[eV]$ does
not exclude a small value (with respect to the elementary
particles scales) for the rest energy of hypothetical
$\Phi$-particle.

5. The {\em parameterized-post-Newtonian(PPN) solution} of
equation (\ref{FEq}) is complicated, but because of the constraint
$p < 10^{-28}$ we may use with great precision Helbig's PPN
formalism \cite{STGExp} (for $\alpha\!=\!{\sfrac 1 3}$). Because
of the condition $\omega\equiv 0$ we obtain much more definite
predictions then usual general relations between $\alpha$ and the
length $l_\Phi$:

$\bullet$ {\em Nordtvedt Effect:}

In MDG a body with significant gravitational self-energy
$E_{{}_G}=\sum_{b\neq c} G{\sfrac {m_b m_c}{|{\em r}_b - {\em
r}_c|} }$ will not move along geodesics due to {\em additional
universal anti-gravitational force}:
\ben
{\em F}_{\!{}_N} = -{\sfrac 2 3} E_{{}_G}\nabla \Phi.
\la{NordtF}
\een
For usual bodies it is too small even at distances $|{\em
r-r}_a|\!\leq\!l_\Phi$, because of the small factor $E_G$. Hence,
in MDG we have no strict strong equivalence principle nevertheless
{\em the week equivalence principle is not violated}.

The experimental data for Nordtvedt effect,  caused by the Sun,
are formulated as a constraint $\eta=0 \pm .0015$  on the
parameter $\eta$ which in MDG becomes a function of the distance
$l$ to the source: $\eta(l)=-{\sfrac 1 2}\left(1+l/l_\Phi\right)
e^{-l/l_\Phi}$. This gives constraint $l_\Phi \leq 2\times 10^{10
}[m]$.

$\bullet$ {\em Time Delay of Electromagnetic Waves}

The action of electromagnetic field does not depend on the field
$\Phi$. Therefore influence of $\Phi$ on the electromagnetic waves
in vacuum is possible only via influence of $\Phi$ on the
space-time metric. The solar system measurements of the time delay
of the electromagnetic pulses give the value $\gamma=1 \pm .001$
of this post Newtonian parameter. In MDG this yields the relation
$(1 \pm .001) g( l_{{}_{AU}})= 1$ and gives once more the
constraint $l_\Phi \leq 2\times 10^{10 }[m]$. Here
$g(l):=1+{\sfrac 1 3 }(1+l/l_\Phi) e^{-l/l_\Phi}$.

$\bullet$ {\em Perihelion Shift}

For the perihelion shift of a planet orbiting around the Sun (with
mass $M_\odot$) in MDG we have: $\delta \varphi = {\frac
{k(l_p)}{g(l_p)}}\delta\varphi_{{}_{GR}}.$ Here $l_p$ is the
semimajor axis of the orbit of planet and $k(l_p) \approx 1 +
{\sfrac 1 {18}}\left( 4 + {\sfrac{l_p^2}{l_\Phi^2}}{\sfrac {l_p
c^2}{GM_\odot}}\right) e^{-l_p/l_\Phi} -{\sfrac 1 {27}}
e^{-2l_p/l_\Phi}$ is obtained neglecting its eccentricity. The
observed value of perihelion shift of Mercury gives the constraint
$l_\Phi \leq  10^{9 }[m]$.

Hence, the known data show that dilaton field $\Phi$ does not
cause observable deviations from GR in solar system. Essential
deviations from Newton law of gravity may not be expected at
distances greater then few $mm$.

\section{Vacuum Energy and True Vacuum Solution in MDG }

Consider the total (true) tensor of energy momentum:
\ben
TT_{\mu\nu}:=
T_{\mu\nu}+ <\rho_0>c^2 g_{\mu\nu},
\la{TT}
\een
$<\!\!\rho_0\!\!>$ being the
averaged energy density of the zero quantum fluctuations. For {\em
true vacuum solution} of MDG: $\Phi\!=\!\Phi_0\!=const,\,\,\,
g_{\mu\nu}=\eta_{\mu\nu}$ (Minkowski metric) from field equations
(\ref{FEq}) we obtain:
\ben
\Phi_0 {\frac{d\Pi}{d\Pi}}(\Phi_0)+\Pi(\Phi_0)=0 \hskip 3.truecm\\
T_{\mu\nu}^0=-{\frac {c^2}{\bar\kappa}}\Lambda U_0 \eta_{\mu\nu}=
TT_{\mu\nu}^0- <\rho_0>c^2 \eta_{\mu\nu},
\la{TVEq}
\een
where $U_0=\Phi_0\Pi(\Phi_0)=\Phi_0\Pi_0$. But for true vacuum
solution we must have $TT_{\mu\nu}^0\equiv 0$ and then \ben
 <\rho_0>=\kappa^{-1}\Lambda U_0=\kappa^{-1}\Lambda
 \Pi_0.
\la{rho}
\een
Hence, true vacuum $(TT_{\mu\nu}=0)$ yields Minkowski space-time,
but physical vacuum $(T_{\mu\nu}=0)$ yields de Sitter space-time.
In our model the real word is de Sitter Universe created by zero
quantum vacuum fluctuations and perturbed by other matter and
radiation fields.

For $<\!\!\rho_0\!\!>$ calculated using Plank length as a quantum
cutoff the observed value of $\Lambda$ gives: \ben\kappa
<\rho_0>/\Lambda = U(\Phi_0)/U(\bar\Phi)\approx
10^{122}\la{122}.\een This huge number yields the famous {\em
cosmological constant problem} and varies from $10^{118}$ to
$10^{123}$ in different articles \cite{Weinberg}. We see that: 1)
It is obviously close in order to the ratio of cosmological action
${\cal A}_c$ and Planck constant $\hbar$:
$U(\Phi_0)/U(\bar\Phi)\approx {\cal A}_c/\hbar$; 2) in MDG there
is no crisis caused by this number, because it gives the ratio of
the values of cosmological potential for different solutions:
$\Phi_0$ and $\bar\Phi$, i.e. {\em for different states of the
Universe}.

If we calculate the values ${\cal A}_{G,\Lambda}^0$ and $\bar{\cal
A}_{G,\Lambda}$ of the very action (\ref{A_Gc}) and introduce
corresponding specific actions per unit volume:
\,$\alpha_0=-\Lambda c \bar\kappa^{-1} U_0$\, and
\,$\bar\alpha=\Lambda c \bar\kappa^{-1}\bar U$,
we can rewrite the above observed result in the form
$\bar\alpha \approx -\alpha_0
\times \hbar /{\cal A}_c= |\alpha_0| \times 10^{-122}$.

We hope that this new formulation of cosmological constant problem
will bring us to its resolution.

It's natural to think that the huge ratio ${\cal A}_c / \hbar
\approx 10^{122}$ is produced during the evolution of the
Universe. To perform qualitative analysis we consider first the
simplest model of Universe build of Bohr hydrogen atoms in ground
state, i.e. we describe the whole content of the Universe using
such {\em effective Bohr hydrogen} (EBH) atoms. Then for the  time
of the existence of Universe $T_U \sim 4\times 10^{17} sec$ one
EBH atom with Bohr angular velocity $\omega_B = m_e
e^4\hbar^{-3}\sim 4\times 10^{16} sec^{-1}$ accumulates classical
action ${\cal A}_{EBH}= 3/2 \,\omega_B T_U\, \hbar \sim 2.4\times
10^{34} \,\hbar$. Hence, the number of EBH in Universe, needed to
explain the present day action $\sim {\cal A}_c$, must be
$N_{EBH}\sim 5\times 10^{87}$. This seems to be quite reasonable
number, taking into account that the observed number of barions in
Universe is $N_{barions}^{obs}\sim 10^{78}$ and we see that in our
approach we have disposable some $9$ orders of magnitude to solve
cosmological constant problem taking into account the contribution
of all other constituents of matter and radiation (quarks ,
leptons, gamma quanta, etc) during the evolution of Universe from
the Big Bang to the present epoch. Neglecting the temperature evolution of
the Universe, we obtain an accumulated action ${\cal A}_\gamma \sim
\omega_\gamma T_U \, \hbar \sim 10^{30}\, \hbar$ for one
$\gamma$-quanta of CMB (which is most significant part of radiation
in Universe). A simple estimate for Bohr-like angular velocity of
constituent quarks in proton is $\omega_{B q}= m_e/m_q (r_B/r_p)^2
\omega_B \approx 10^7 \omega_B$ (for mass of constituent quark
$m_q \sim 5 MeV$, Bohr radius $r_B$ and radius of proton $r_p\sim
8 \times 10^{-13}cm $). Then the action, accumulated by constituent
quarks in one proton during evolution of the Universe, is ${\cal
A}_{p}\sim \omega_{B q} T_U\, \hbar \sim 10^{42} \,\hbar$. This
gives unexpectedly good estimation for the number of effective
protons (ep) in the Universe: $N_{ep}\sim 10^{80}$. We may use the
left-off two orders of magnitude to take into account contribution
of the other matter constituents and of the temperature evolution of
the Universe: during the short-time initial hot phase some additional
action must be produced.

The main conclusion of this qualitative consideration based on
classical mechanics and simplest application of basic quantum
relations is that actually in MDG {\em the observed nonzero value}
of cosmological constant $\Lambda^{obs}\neq 0$ {\em restricts the
number of degrees of freedom} in the observable Universe and maybe
forbids the existence of  more deep levels of matter below
the quark level.

\section{Application of MDG  in Cosmology}

In MDG for Freedman-Robertson-Walker (FRW) adiabatic homogeneous
isotropic Universe with $ds^2_{FRW}= c^2 dt^2 - A^2 dl^2_k$,\,
$t=T_c\tau$,\, $A(t)=A_c a(\tau)$ and dimensionless $dl^2_k={\frac
{dl^2}{1-kl^2}+l^2(d\theta^2+sin^2\theta)d\varphi^2}$\,($k=-1, 0,
1$) in presence of matter with energy-density
$\varepsilon=\varepsilon_c\epsilon(a)/\bar\Phi$ and pressure
$p=\varepsilon_c p_\epsilon(a)/\bar\Phi$ dynamical equations are:
\ben
{\sfrac 1 a}{\sfrac {d^2\!a}{d\tau^2}}+
{\sfrac 1 {a^2}}({\sfrac {da}{d\tau}})^{{}_2}+{\sfrac k {a^2}}=
{\sfrac1 3} \left(\Phi{\sfrac{d\Pi}{d\Phi}}(\Phi)+\Pi(\Phi)\right),\nonumber\\
{\sfrac 1 a}{\sfrac {da}{d\tau}}{\sfrac {d\Phi}{d\tau}}+
\Phi \left({\sfrac 1 {a^2}}({\sfrac {da}{d\tau}})^{{}_2}
+{\sfrac k {a^2}}\right)={\sfrac1 3}\left(\Phi\Pi(\Phi)+ \epsilon(a)\right).
\la{DERWU}
\een

The use of Hubble parameter $h(a)=a^{-1}{\sfrac
{da}{d\tau}}(\tau(a))$ (where $\tau(a)$ is the inverse function of
$a(\tau)$), new variable $\lambda=\ln a$ and prime for
differentiation with respect to $\lambda$ gives the system for
$\Phi(\lambda)$ and $h^2(\lambda)$:
\ben
{\sfrac 1 2}(h^2)^\prime +2 h^2 + k e^{-2\lambda}= {\sfrac 1
3}\left(\Phi{\sfrac{d\Pi}{d\Phi}}(\Phi)+\Pi(\Phi)\right),\nonumber
\\ h^2 \Phi^\prime +\left(h^2+k e^{-2\lambda}\right)\Phi= {\sfrac
1 3}(\Phi\Pi(\Phi)+\epsilon(e^\lambda))\nonumber
\la{NDE}
\een
and relation $\tau(a)=\int^a_{\!a_{in}}\!da /(a\,h(a))+\tau_{in}$.
Excluding cosmological factor $\Pi(\Phi)$ we obtain the equation:
\ben
\Phi^{\prime\prime}\!+\! \left({\sfrac {h^{\prime}}
h}\!-\!1\right)\Phi^{\prime} \!+\!2\left({\sfrac {h^{\prime}}
h}\!- k h^{-2}e^{-2\lambda}\right)\Phi \!=\!{\sfrac 1
{3h^2}}\epsilon^{\prime},
\la{ODEPhi}
\een
or in terms of the function $\psi(a) = \sqrt{|h(a)|/a}\,\Phi(a)$:
\ben
\psi^{\prime\prime} + n^2 \psi = \delta,
\la{DEPsi}
\een
$-n^2 = {\sfrac 1 2}{\sfrac {h^{\prime\prime}} h}- {\sfrac 1
4}({\sfrac {h^{\prime}} h})^2\!- {\sfrac 5 2}{\sfrac {h^{\prime}}
h}+{\sfrac 1 4} +{\sfrac {2 k} {h^2}} e^{-2\lambda}$,\,$ \delta =
{\sfrac 1 3}\sqrt{{ a/{|h|^3}}} {\sfrac {d\epsilon}{da}}$.

Now we are ready to consider {\em the inverse cosmological
problem}: to find a cosmological factor $\Pi(\Phi)$ which yield
given evolution $a(\tau)$ of the Universe. A remarkable property
of MDG is that unique solution of this problem exist for almost
any three times differentiable function $a(\tau)$; the values of
all "bar" quantities (including $\bar \kappa$ in action
(\ref{A_Gc})) may be determined from time evolution $a(\tau)$ of
the Universe via the solution $\bar \lambda = \ln \bar a$ of the
Eq. (\ref{RV}).

Indeed: for known $a(\tau)$ construct the function $h(\lambda)$
and find the point $\bar\lambda$ as real solution of the algebraic
equation $r(\bar\lambda)\!=\!-4$,\,
$r(\lambda)\!=\!-6\left({\sfrac 1 2}(h^2)^\prime\!+\!2 h^2\!+\!k
e^{-2\lambda}\right)$ being dimensionless scalar curvature: $r=
R/\Lambda$. Then using Eq. (\ref{CFC}) obtain
$\bar\Phi\!=\!-4\bar\epsilon \left(1\!+\!{\sfrac 4 3}p^2\right)/
\!\left(\bar j_{00}^\prime(1\!+\!{\sfrac 4 3}p^2) + 4p^2\bar
h^2\bar r^\prime \right)$, $\bar\Phi^\prime/\bar \Phi\!=\!-{\sfrac
1 3}p^2 \bar r^\prime/\left(1\!+\!{\sfrac 4 3}p^2\right)$;\,
$j_{00}\!\!=\!G_{00}/\Lambda\!=\!3\left(h^2\!+\!ke^{-2\lambda}\right)$
is dimensionless $00$-component of Einstein tensor. In their turn
quantities $\bar\Phi$ and $\bar\Phi^\prime$ determine the values
of constants $C_{1,2}$ in general solution $\Phi(\lambda)$ of Eq.
(\ref{ODEPhi}):
$\Phi(\lambda)=C_1\Phi_1(\lambda)+C_2\Phi_2(\lambda)+
\Phi_\epsilon(\lambda)$\,where $\Phi_1(\lambda)$ and
$\Phi_2(\lambda)$ are a fundamental system of solutions of
corresponding homogeneous equation and
$$\Phi_\epsilon\!=\!{\frac{\bar a}{(3\bar h\bar\Delta)}}\!\left(
\Phi_2\!\int_{\bar\lambda}^\lambda\!d\epsilon\,{\frac
{\Phi_1}{ah}}- \Phi_1\!\int_{\bar\lambda}^\lambda\!d\epsilon\,
{\frac {\Phi_2} {ah} } \right),$$ $\Delta(\lambda)=
\Phi_1\Phi_2^\prime-\Phi_2\Phi_1^\prime$. The dependence on
$\lambda$ of cosmological factor $\Pi$ and potential $V$ are given
by equations \ben \Pi(\lambda) = j_{00} + 3 h^2 \Phi^\prime/\Phi -
\epsilon/\Phi, \nonumber \\ V(\lambda)={\sfrac 2 3}
\int\Phi\left(\Phi\Pi^\prime-\Phi^\prime\Pi \right)d\lambda
\la{PiV} \een which define functions $\Pi(\Phi)$ and $V(\Phi)$
implicitly.

This mathematical result shows maybe the best way to study SUSY
breaking and the corresponding potential $V_{SUSY}(\phi)$: we have
to reconstruct the {\em real time evolution} $a(\tau)$ of the
Universe from astrophysical observations.

Finally we stress following specific properties of MDG:

1) If $n\!>\!0$ {\em dilatonic field $\Phi(a)$ oscillates}; if
$n\!<\!0$ such oscillations do not exist. The change of sign of
dilaton field $\Phi$ yields phase transitions of Universe from
gravity ($\Phi\!>\!0$) to anti-gravity ($\Phi\!<\!0$), or
vice-versa which are possible for width class of cosmological
potentials, but excluded for other potentials.

2) In spirit of Max principle Newton constant depends on presence
of matter: $G\!\sim\!1/\bar\Phi\!\sim\!1/\bar\epsilon$.

3) For simple functions $a(\tau)$ the cosmological factor
$\Pi(\Phi)$ and potentials $V(\Phi)$ and $U(\Phi)$ may show
unexpected {\em catastrophic behavior:} $\sim\!(\Delta\Phi)^{3/2}$
($\Delta\Phi\!=\!\Phi\!-\!\Phi(\lambda^\star)$) in vicinity of the
critical points $\lambda^\star$:
$\Phi^\prime(\lambda^\star)\!=\!0$ of the projection of analytical
curve $\{\Pi(\lambda),\Phi(\lambda),\lambda\}$ on the plain
$\{\Pi,\Phi\}$. Scale factors $a(\tau)$ yielding an {\em
everywhere analytical cosmological factor} $\Pi(\Phi)$ exist, too.

4) Clearly one can construct  MDG model of Universe {\em without
initial singularities}: $a(\tau_0)=0$ (which are typical for GR)
and with {\em any desired kind of inflation.}

5) Because the dilaton field $\Phi$ is quite massive, in it will
be stored significant amount of energy. An interesting open
question is:  may the field $\Phi$ play the role of dark matter in
the Universe?

A very important problem is to reconstruct the cosmological factor
$\Pi(\Phi)$ of real Universe using proper experimental data and
astrophysical observations. This problem was at first studied in
\cite{Starobinsky} for more complicated models than MDG with two
independent unknown functions.

We see that MDG is a rich model with new curious features and
deserves further careful investigation.

\vskip 1truecm

{\em Acknowledgments:} The author is deeply indebted to
A.~D.~Dolgov and L.~B.~Okun' for discussions on some properties of MDG
and its connection with cosmological constant problem,
to G.~Esposito-Far\'ese for the useful additional information about
the scalar-tensor theories of gravity in connection with MDG,
to A.~A.~Starobinsky for his kind help in the
references and for his comment about the relation of MDG model and
Starobinsky model \cite{Starobinsky80} and
to E.~Gozzi for his help and kind hospitality in the Department of
Theoretical Physics of the University of Trieste, where a part of
this work has been performed.

\end{document}